\documentclass{pasj00}

\begin{document}
\SetRunningHead{S.Nakashima et al.}{Suzaku Observations of the Great Annihilator and the Surroundings}
\Received{2010/04/05}
\Accepted{2010/04/30}

\title{Suzaku Observations of the Great Annihilator \\ and the Surrounding Diffuse Emissions}

\author{Shinya \textsc{Nakashima}, Masayoshi \textsc{Nobukawa}, Takeshi Go \textsc{Tsuru}, Katsuji \textsc{Koyama}} 
\affil{Department of Physics, Graduate School of Science, Kyoto University, Sakyo-ku, Kyoto 606-8502}
\email{shinya@cr.scphys.kyoto-u.ac.jp}
\and
\author{Hideki \textsc{Uchiyama}} 
\affil{Department of Physics, The University of Tokyo, 7-3-1 Hongo, Bunkyo-ku, Tokyo 113-0033}

%

\KeyWords{Galaxy: center---ISM: clouds---ISM: supernova remnants---X-rays: individual (Great Annihilator)} 

\maketitle

\begin{abstract}
We report the Suzaku observation of 1E~1740.7$-$2942, 
a black hole candidate called the ``Great Annihilator'' (GA).
The high-quality spectrum of Suzaku provides the severest constraints 
on the parameters of the GA.  
Two clumpy structures are found around the GA in the line images of 
Fe\emissiontype{I} K$\alpha$ at 6.4~keV and S\emissiontype{XV} K$\alpha$ at 2.45~keV.
One clump named M\,359.23$-$0.04 exhibits the 6.4-keV line with an equivalent width of
$\sim$ 1.2~keV, and is associated with a molecular cloud in the radio CS($J$=1--0) map. 
Thus the 6.4-keV line from M\,359.23$-$0.04 is likely due to X-ray fluorescence 
irradiated by an external X-ray source.
The irradiating X-rays would be either the past flare of Sagittarius\,A$^*$ or 
the bright nearby source, the GA.
The other clump named G\,359.12$-$0.05 is associated with the radio supernova remnant candidate 
G\,359.07$-$0.02.
We therefore propose that G\,359.12$-$0.05 is an X-ray counterpart of G\,359.07$-$0.02.  
G\,359.12$-$0.05 has a thin thermal plasma spectrum with a temperature of $kT \sim 0.9$~keV. 
The plasma parameters of G\,359.12$-$0.05 are consistent with those of 
a single supernova remnant in the Galactic center region.
\end{abstract}

\section{Introduction}
The Galactic center (GC) region has many celestial objects such as a supermassive black hole Sagittarius\,A$^*$ (Sgr\,A$^*$),
star-forming regions, dense molecular clouds (MCs).
The GC region is also full of high-energy phenomena.
For example, several supernova remnants (SNRs) or candidates have been recently found (e.g., \cite{Nobukawa.2008}; \cite{Mori.2008}; \cite{Sawada.2009}; \cite{Tsuru.2009}). 
The most characteristic phenomenon is the diffuse 6.7-keV lines from highly ionized irons (\cite{Koyama.1989}; \cite{Yamauchi.1990}).
The origin of this emission is probably due to a plasma with $kT~\sim$~6.5 keV temperature \citep{Koyama.2007b}, 
but is still a debatable issue.
Another notable feature in the GC region is the clumpy 6.4-keV lines from neutral irons
(e.g., \cite{Koyama.1996}).
The origin is considered as MCs illuminated by X-rays or electrons.
Some of the 6.4-keV clumps are well explained by the irradiation of the past X-ray flare from Sgr\,A$^*$ (e.g., \cite{Nobukawa.2008}; \cite{Inui.2009}; \cite{Nakajima.2009}).

1E~1740.7$-$2942 was discovered by the Einstein observatory \citep{Hertz.1984} in the direction of the GC region, 
and was found to be the brightest GC source at above 20~keV (\cite{Skinner.1987}; \cite{Sunyaev.1991}).
The time variation of the flux and the spectrum of 1E~1740.7$-$2942 are similar to those of Cygnus\,X-1, the archetypal 
black hole candidate, and hence 1E~1740.7$-$2942 has been considered to be a black hole candidate (\cite{Cook.1991}; \cite{Skinner.1991}).   

GRANAT detected a prominent bump on the spectrum of 1E~1740.7$-$2942 at 300--600 keV, which was interpreted  
as an electron-positron annihilation line at 511~keV (\cite{Bouchet.1991}; \cite{Sunyaev.1991}; \cite{Churazov.1993a}; \cite{Cordier.1993}). 
Thus 1E~1740.7$-$2942 is named the ``Great Annihilator'' (GA).
However, no evidence for the annihilation line has been found so far by other satellites 
(\cite{Harris.1994}; \cite{Jung.1995}; \cite{Smith.1996}; \cite{Cheng.1998}; \cite{Bouchet.2009}).
 
The VLA radio observations discovered a radio counterpart of the GA and non-thermal double jet-like 
structures emanating from the GA \citep{Mirabel.1992}, and hence the GA is a ``micro quasar''. 
Like the other Galactic jet sources, the GA would be a binary system with a stellar-mass black hole. 
In fact, the possible orbital period of 12.73 days was discovered by \citet{Smith.2002}.
However, no clear companion star was found with the optical, infrared or radio observations, due mainly to 
the strong interstellar absorption toward the GC region (e.g., \cite{Eikenberry.2001}).

The X-ray spectrum of the GA below 10 keV is explained by an absorbed power-law model (\cite{Sakano.1999}; \cite{Gallo.2002}).
\citet{Cui.2001} reported the extended X-ray emission which is perpendicular to the radio-jets (but see \cite{Gallo.2002}).  
On the other hand, the INTEGRAL observations suggested that the spectrum in the
10--100~keV band is explained by a model of either a comptonized-plasma plus a power-law 
or two comptonized-plasmas \citep{Bouchet.2009}.

The radio observations found a giant MC near the GA (\cite{Bally.1991}; \cite{Mirabel.1991}).  
Then the authors proposed the Bondi-Hoyle accretion mechanism \citep{Bondi.1944}; the GA is in the MC, and is powered
by the gas accretion from the MC.  
However the ASCA observations found no clear evidence for dense gas around the GA (\cite{Churazov.1996}; \cite{Sakano.1999}).

In order to investigate the accretion mechanisms, whether the Bondi-Hoyle process, 
due to a binary companion or else, we performed the Suzaku observation on the GA, 
with its high-quality spectrum and low background.  The other objective of the observation
is to  discover local structures, if any, on the GA and in the close vicinity of the GA,
and investigate the physical relation to the GA.
We assume the distance to the GC to be 8.5~kpc in this paper.

\section{Observations and Data Reduction}

\begin{table*}[t]
\begin{center}
\caption{Observation log}\label{table1}
 \begin{tabular}{cccccc}
  \hline
  Target name & Observation ID & \multicolumn{2}{c}{Pointing direction} & Observation date & Exposure \\ 
              &        &  $\alpha$ (J2000.0)  &  $\delta$ (J2000.0) &         &  (ks)    \\
  \hline
  GC\_LARGEPROGECT4 & 503010010 & \timeform{17h44m10s} & \timeform{-29D33'20''} & 2008-09-06 & 53.1 \\
  GC\_LARGEPROGECT5 & 503011010 & \timeform{17h43m47s} & \timeform{-29D49'59''} & 2008-09-08 & 57.6 \\
  \hline
 \end{tabular}
\end{center}
\end{table*} 

Two pointing observations toward the GA were performed in September 2008 with the X-ray Imaging Spectrometer (XIS: \cite{Koyama.2007a}) at the focal plane of the X-Ray Telescope (XRT: \cite{Serlemitsos.2007}) on board the Suzaku satellite (\cite{Mitsuda.2007}). The observation log is given in table \ref{table1}.

The XIS system consists of three sets of front-illuminated (FI) CCD cameras (XIS\,0, 2 and 3) and one set of a back-illuminated (BI) CCD camera (XIS\,1).
The performance of the CCD cameras has been gradually degraded due to the radiation damage by cosmic-ray particles, thus the Spaced-row Charge Injection (SCI) technique was introduced to restore the energy resolution since October 2006 \citep{Uchiyama.2009}.
Thanks to the SCI, XIS energy resolutions (FI/BI) at 5.9~keV were 155/175~eV (FWHM) during the observations.
However, XIS\,2 is unusable for scientific observations because of the sudden anomaly in 2006, and we use the data from 
the remaining three cameras.

The XRT consists of closely nested thin-foil reflectors and has large collecting efficiency (450~cm$^2$ at 1.5~keV and 250~cm$^2$ at 7~keV per XRT).
The point-spread function (PSF) of the XRT is about \timeform{2'} in a half-power diameter.
Each CCD camera has the \timeform{17'.8} $\times$ \timeform{17'.8} field of view. The band pass of the XIS+XRT is  0.3--12~keV.

The observations were carried out using the normal clocking mode with no window/burst option.
The data were screened with the processing version 
2.2.11.22\footnote{http://www.astro.isas.ac.jp/suzaku/process/history/v221122.html} to 
exclude the events taken during passages of the 
South Atlantic Anomaly, the Night-earth elevation angle $<$ \timeform{5D}, and the Day-earth elevation angle $<$ \timeform{20D}.
The effective exposures of the screened data are listed in table \ref{table1}.

\section{Analysis and Results}

The analysis is made with HEASoft version 6.7\footnote{http://heasarc.gsfc.nasa.gov/docs/software/lheasoft} 
and XSPEC version 12.5.1\footnote{http://heasarc.gsfc.nasa.gov/docs/xanadu/xspec}.
The non--X-ray background (NXB) is constructed using {\tt xisnxbgen}, and is subtracted from the data.
For the model fitting of the spectrum, we generate redistribution matrix files and auxiliary response functions 
using {\tt xisrmfgen} and {\tt xissimarfgen}, respectively.
We fit the FI and BI spectra simultaneously, where the FI data of  XIS\,0 and 3 are combined.

\subsection{Wide Band X-ray Image}

\begin{figure}[t]
  \begin{center}
    \FigureFile(80mm,80mm){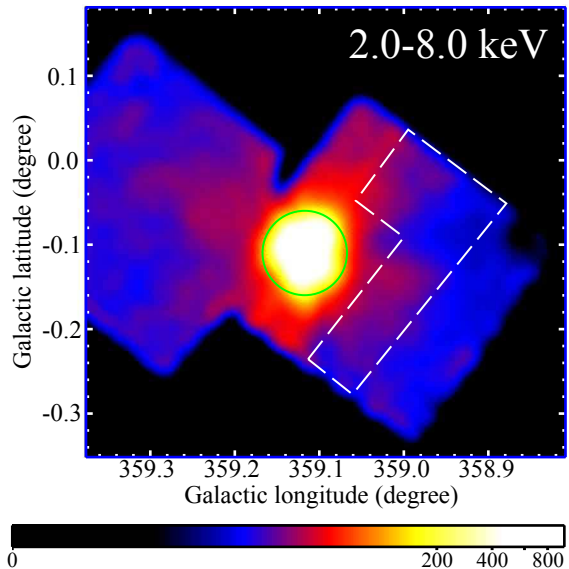}
  \end{center}
  \caption{The XIS image in the 2.0--8.0~keV band smoothed with a Gaussian kernel of $\sigma$ = \timeform{1'}. 
The NXB is subtracted and the vignetting effect is corrected.
The green solid line indicates the source region of the GA, while the white dashed line indicates the background region of the GA.
The color scale is logarithmic in units of counts bin$^{-1}$ (the bin size is \timeform{12.5''} $\times$ \timeform{12.5''}).}\label{figure1}
\end{figure}

In order to see the overall structure, we show the 2.0--8.0~keV band image in figure \ref{figure1},
where all the CCDs data are merged and the calibration source regions are removed.
For visibility, we sum 12 $\times$ 12 pixels (\timeform{12.5''} $\times$ \timeform{12.5''}) and convolve with a Gaussian kernel of $\sigma$ = \timeform{1'}. 
In figure \ref{figure1}, we find a bright source at ($\alpha$, $\delta$)$_{\mathrm{J2000.0}}$ = 
(\timeform{265D.9829}, \timeform{-29D.7462}) of the Suzaku coordinates.
Since this position coincides with the GA position of ($\alpha$, $\delta$)$_{\mathrm{J2000.0}}$ = (\timeform{265D.9758}, 
\timeform{-29D.7501}) \citep{Muno.2006} within the Suzaku 
nominal error of \timeform{19''} \citep{Uchiyama.2008}, we identify this source as the GA.
We then fine-tune the Suzaku coordinates by shifting $\Delta$($\alpha$, $\delta$) = (\timeform{-0D.0071}, \timeform{-0D.0039}).   

\begin{figure}[t]
\begin{center}
    \FigureFile(80mm,55mm){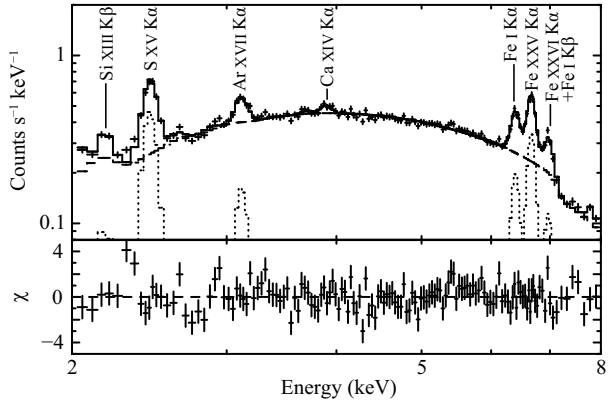}
  \end{center}
  \caption{The XIS FI spectrum extracted from the entire field shown in figure \ref{figure1} without the \timeform{3'}-radius circle centered at 
the GA and calibration source regions. The NXB is subtracted. The dashed line is the best-fit continuum 
(absorbed power-law) and the dotted lines are 
those of the Gaussian lines (see text).}\label{figure2}
\end{figure}

In figure \ref{figure1}, we also see local structures around the GA, partly due to the tail (spill-over)
of the PSF of the XRT, but some are real X-ray structures. The major X-ray emissions around 
the GA are the Galactic Center Diffuse X-rays (GCDX: \cite{Koyama.1996}).
The GCDX is characterized by the strong lines of Fe\emissiontype{I} K$\alpha$, S\emissiontype{XV} K$\alpha$, 
and Fe\emissiontype{XXV} K$\alpha$ at the respective energies of 6.4~keV, 2.45~keV and 6.7~keV.

In order to investigate the fluxes of the prominent lines and their spatial variations in the observed field, 
we make the 2.0--8.0~keV band spectrum from the entire field shown in figure \ref{figure1} except the \timeform{3'}-radius circle 
centered at the GA and the calibration source regions. The spectrum is shown in figure \ref{figure2}. 
We fit the spectrum with a phenomenological model of an absorbed power-law continuum plus Gaussian lines.
We hereafter use the cross sections from \citet{Balucinska-Church.1992} with the solar 
abundance (\cite{Anders.1989}).
Then we see prominent lines at 6.4~keV, 2.45~keV and 6.7~keV above the best-fit continuum 
(dashed line in figure \ref{figure2}).

The 6.4-keV line comes from MCs and hence has clumpy spatial distribution (the 6.4-keV clumps, e.g., \cite{Koyama.1996}; \cite{Murakami.2000}; \cite{Ryu.2009}),
while the 6.7-keV line comes from the \textit{kT}~$\sim$~6.5~keV plasma (the 6.5-keV plasma) 
and is reported to be  more uniform. 
The 2.45-keV line is due to another hot plasma with the temperature of \textit{kT}~$\sim$~1~keV(the 1-keV plasma), 
which is not as uniform as the 6.5-keV plasma but is less clumpy than the 6.4-keV line \citep{Ryu.2009}.
Thus either the 6.4-keV line or the 1-keV plasma may contribute to the local structures.

\subsection{Line Images}

In order to depict the local structures, we use the 2.45-keV, 6.4-keV and 6.7-keV lines. 
To make the 2.45-keV line image, we subtract the continuum flux in the 2.7--3.0~keV (continuum-dominant band) image
from the 2.35--2.55~keV (line-dominant band) image by normalizing the continuum flux ratio
to that estimated from the best-fit continuum (absorbed power-law, see figure \ref{figure2}).  

\begin{figure*}[t]
  \begin{center}
    \FigureFile(160mm,163mm){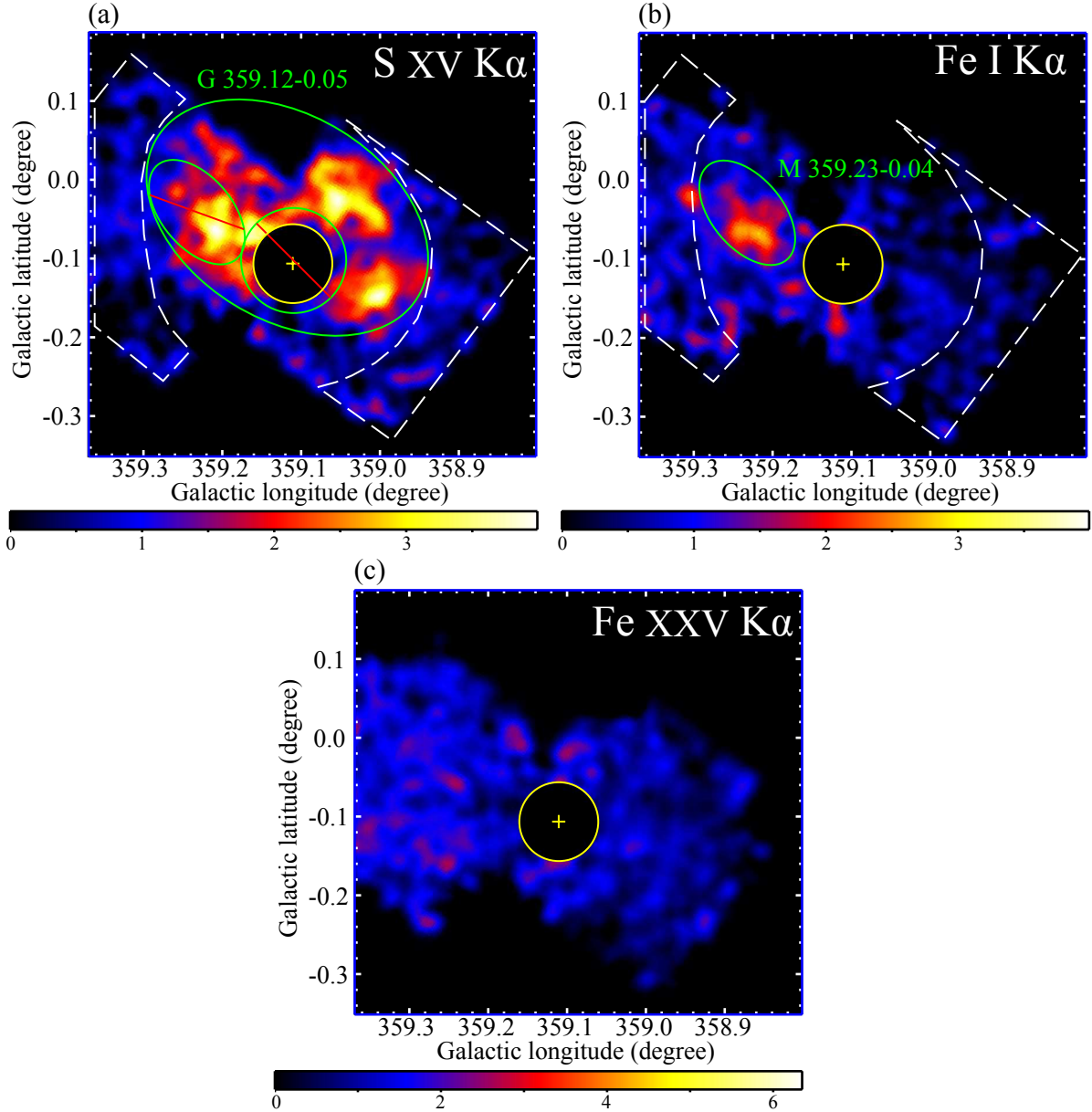}
  \end{center}
  \caption{The continuum-subtracted (see text) images around the GA (the yellow cross) in the S\emissiontype{XV} K$\alpha$ line (a), Fe\emissiontype{I} K$\alpha$ line (b), and Fe\emissiontype{XXV} K$\alpha$ line (c), 
where the \timeform{3'}-radius regions (the yellow circle) centered at the GA are excluded.
The vignetting effect is corrected, and the images are smoothed with a Gaussian kernel of $\sigma$ = \timeform{1'}. 
The green solid lines and the white dashed lines indicate the source and the background regions, respectively. 
The region of G\,359.12$-$0.05 is a \timeform{16'} $\times$ \timeform{24'} ellipse excluding the circle of \timeform{4'}-radius 
centered at the GA and the ellipse of the other source M\,359.23$-$0.04 (shown in (a)). The region of M\,359.23$-$0.04 is a \timeform{5'} $\times$ \timeform{9'} ellipse (shown in (b)). The color scales are liner in units of 10$^{-5}$ counts s$^{-1}$ bin$^{-1}$ (the bin size is \timeform{12.5''} $\times$ \timeform{12.5''}).}\label{figure3}
\end{figure*}

As for the 6.4-keV and 6.7-keV line images, we use the line-dominant bands of 6.3--6.5~keV and 6.6--6.8~keV,
respectively, and subtract the continuum-dominant flux in the 5.0--6.0~keV band with the same method as 
the 2.45-keV line image.
The continuum-subtracted line images are shown in figure \ref{figure3}, 
where the \timeform{3'}-radius circle centered at the GA is excluded to emphasize the surrounding structures. 
We correct the vignetting effect and smooth with a Gaussian kernel of $\sigma$ = \timeform{1'}.

In the 2.45-keV line image (figure \ref{figure3}a), we find a diffuse structure (\timeform{16'} $\times$ \timeform{24'}) surrounding the GA, 
which is larger than the X-ray halo (\timeform{40''}) reported by \citet{Gallo.2002}. From the center position, we name this structure G\,359.12$-$0.05.
In the 6.4-keV line image (figure \ref{figure3}b), we find a clump in the north (left-side of figure \ref{figure3}b) of the GA, 
and name the clump M\,359.23$-$0.04.
We note that no local structure in the 6.7-keV line image (figure \ref{figure3}c) is found.
This supports that the 6.7-keV line (the 6.5-keV plasma) is uniform, at least near and around the GA.

\subsection{Light Curve and Spectrum of the Great Annihilator}

A light curve in the 2.0--8.0~keV band is made from the \timeform{3'}-radius circle centered at the GA (figure \ref{figure1}).  
No significant periodicity is found in the light curve in the time range of
8--10000 sec. Also, no flux-variation larger than 10\% 
is found in the time scale longer than 512 sec.

\begin{figure}[t]
  \begin{center}
    \FigureFile(80mm,55mm){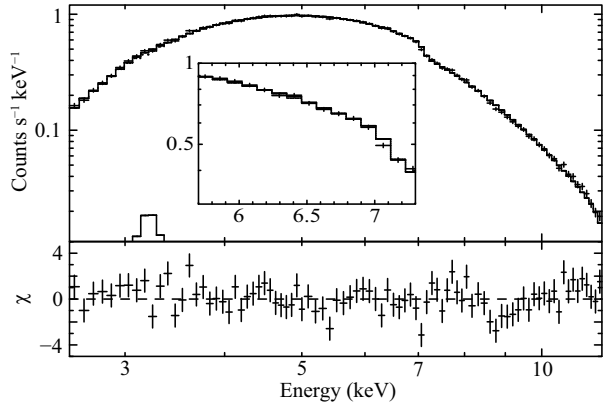}
  \end{center}
  \caption{The XIS FI background-subtracted spectrum of the GA in the 2.5--12~keV band with the logarithmic scales
for both the horizontal and vertical axes.
The enlarged spectrum in the 5.7--7.3~keV band is given in the inset, where the horizontal axis is liner
and the vertical axis is logarithmic. We see no line-like structure at 6.4 keV (see text).}\label{figure4}
\end{figure}

We therefore sum the data in all the observations and make the X-ray spectrum 
from the source region (the solid circle in figure \ref{figure1}).
The background spectrum made from the dashed box in figure \ref{figure1} is subtracted 
after correcting the difference of the effective area.
For the correction of the effective area, we take into account the energy dependence of the effective area
with the method given in \citet{Hyodo.2008}. 
We then fit the 2.5--12.0~keV band spectrum with an absorbed power-law model.  
Since a clear iron absorption edge at 7.1 keV is found, the iron abundance for the photoelectric absorption 
is treated as a free parameter, although the abundances of the other elements are fixed to the solar values.
We find an excess bump at 3.2~keV, which may be the calibration 
error of the 
XRT Au M-II edge (\cite{Kubota.2007}). Thus we add a Gaussian at this energy with the 
fixed line-width of 0.1~keV (1 $\sigma$).
Since the 6.4-keV line carries important information for the circumstellar gas around the GA (e.g., \cite{Churazov.1996}; \cite{Sakano.1999}),  we add a Gaussian line at 6.4~keV and fit the spectrum.  
Then we constrain the upper-limit of the equivalent width at 6.4~keV (\textit{EW}$_\mathrm{6.4\,keV}$) to be $<$ 7.8~eV at the 90\% confidence level.
The best-fit spectrum  and the best-fit parameters are shown in figure \ref{figure4} and in table \ref{table2}, respectively.
Note that we show only the FI spectrum in figure \ref{figure4} for simplicity, although the FI and BI spectra are fitted simultaneously.

The error due to possible flux variation in the background spectrum  would be small, because the background spectrum 
is made from the region where the 6.4-keV-line flux is smoothly distributed (see figure \ref{figure3}b). 
For a more quantitative estimate, we drive the 6.4-keV-line flux 
in every \timeform{3'}-radius segment in the entire region of figure \ref{figure3}b, excluding the areas of the GA and M\,359.23$-$0.04. 
Then we find that the flux-variation of the 6.4-keV line is within $\sim~1.1 \times 10^{-3}$~counts~s$^{-1}$ 
(90\% confidence).  This value is about 1/5 of the upper limit of
the 6.4-keV-line flux from the GA ($\sim~5.8 \times 10^{-3}~\mathrm{counts}~\mathrm{s}^{-1}$). 
Adding this error due to the background fluctuation to the statistical error, we find that the upper-limit of \textit{EW}$_\mathrm{6.4\,keV}$ is 8 eV.

\begin{table}[t]
\begin{center}
\caption{Best-fit parameters of the GA}\label{table2}
 \begin{tabular}{ll}
  \hline
  \multicolumn{1}{c}{Parameter} & \multicolumn{1}{c}{Value} \\
  \hline
  $N_{\rm{H}}$ (10$^{22}$ cm$^{-2}$) & 12.2 (11.9--12.6) \\
  $Z_{\rm{Fe}}$\footnotemark[$^*$] & 1.35 (1.23--1.47) \\
  Photon index & 1.41 (1.38--1.44) \\
  Flux\footnotemark[$\dagger$] (photons cm$^{-2}$ s$^{-1}$) & 1.73 $\times$ 10$^{-2}$ \\
  Luminosity\footnotemark[$\ddagger$] (erg s$^{-1}$) & 2.60 $\times$ 10$^{36}$ \\
  6.4-keV line:\\
  \phantom{0}Flux (counts s$^{-1}$) & $<$ 5.8 $\times$ 10$^{-3}$ \\
  \phantom{0}\textit{EW}$_\mathrm{6.4~keV}$ (eV) & $<$ 7.8 \\
  $\chi^2$/\textit{d.o.f} & 243/170 \\
  \hline
  \multicolumn{2}{@{}l@{}}{\hbox to 0pt{\parbox{75mm}{\footnotesize
	Notes. The errors are at the 90\% confidence level.
	\par\noindent
	\footnotemark[$^*$] Iron abundance in units of solar. The abundances of the other elements are fixed to the solar values.
	\par\noindent
	\footnotemark[$\dagger$] Observed flux in the 2.5--12.0~keV band. 
	\par\noindent
	\footnotemark[$\ddagger$] Absorption-corrected luminosity in the 2.5--12.0~keV band at the distance of 8.5~kpc.
  }\hss}}
 \end{tabular}
\end{center}
\end{table} 

\subsection{Spectrum of G\,359.12$-$0.05}

\begin{figure}[t]
  \begin{center}
    \FigureFile(80mm,60mm){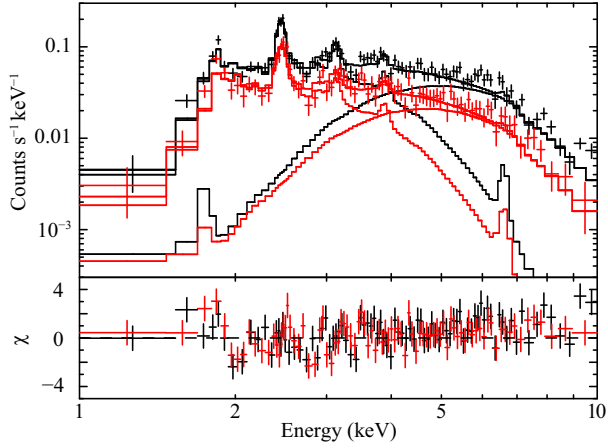}
  \end{center}
  \caption{The background subtracted spectra of G\,359.12$-$0.05 for the FI CCD (black) and
BI CCD (red). The solid lines show the models of the absorbed plasma emission and the spill-over 
component of the GA.}\label{figure5}
\end{figure}

We extract the spectrum of G~359.12$-$0.05 from the source region of the solid ellipse in figure \ref{figure3}a,
excluding the circular region of \timeform{4'}-radius around the GA and
the elliptical region of the other source M\,359.23$-$0.04 (figure \ref{figure3}b). 
The background spectrum is made from the two regions with no local enhancement in the 
2.45-keV, 6.4-keV and 6.7-keV line images (the dashed areas).
The background-subtracted spectra are shown in figure \ref{figure5}.
We find that the spectrum clearly extends to the hard band and that it shows several emission lines from highly ionized atoms in the soft band. 

The line structures in the soft band are due to a hot thin plasma, while the main origin of 
the hard band X-rays is spill-over of the GA flux due to the tail of the PSF of the XRT.
Using the best-fit spectra of the GA (table 2), we simulate the spill-over flux and spectrum by the 
ray-tracing software ({\tt xissimarfgen}). 
We then fit the spectra with the model of an absorbed thin thermal plasma (VAPEC in XSPEC) plus the spill-over 
component.
The best-fit spectrum and parameters are shown in figure \ref{figure5}  and table \ref{table3}, respectively.

Statistically, the fit is rejected with $\chi^2$/\textit{d.o.f} = 1.54 (\textit{d.o.f} = 166). 
In fact, we see a hint of residual excess in the hard band.
Two possibilities may be considered for this excess.
The first is the under-subtraction of the GCDX plasma of \textit{kT} $\sim$ 6.5~keV, 
due to possible non-uniformity near the GA. 
This possibility may be negligible, because we neither see a local enhancement in the 6.7~keV line image (figure \ref{figure3}c) nor a hint of the 6.7-keV line in the G\,359.12$-$0.05 spectrum.
Nevertheless, we add a 6.5-keV plasma to the fitting-model, and find that the possible 
contamination of the GCDX is less than 15\% in the 1--10 keV band.  

The second possibility is a calibration error of the spill-over component.
We therefore allow the spill-over flux to be a free parameter. 
The fit is improved to $\chi^2$/\textit{d.o.f} = 1.09 (\textit{d.o.f} = 164). 
The additional excesses in the hard band are respectively 31\% (FI) and 23\% (BI) of 
those estimated with the nominal PSF calibration.

The nominal PSF calibration uncertainties\footnote{http://www.astro.isas.ac.jp/suzaku/doc/suzakumemo/suzakumemo-2008-04.pdf} 
beyond \timeform{2'} are  5\% and 15\% in FI and BI, respectively.
We note that the PSF calibration was performed in the on-axis angle and 
concentric circle regions.
Our observations were, on the other hand, performed in the off-axis angle ($\sim$ \timeform{6'})
and the source region is non-concentric around the GA.
We hence compare the simulated flux \textit{vs} observed flux using the Seyfert galaxy MCG\,$-$5--23--16,
which is the \timeform{3.5'} off-axis angle observation \citep{Reeves.2007}.
Unlike the GA in the GCDX, MCG\,$-$5--23--16 is located off the Galactic-plane, where the major X-ray background is 
the cosmic--X-ray background (CXB). The CXB is spatially more uniform than the GCDX. Therefore 
we can obtain more reliable fluxes on and around the source.
We also adopt the non-concentric region from the MCG\,$-$5--23--16, the similar region of 
G\,359.12$-$0.05. Then we compare the flux of the observed data 
(after the subtraction of the NXB and CXB) 
and that of the ray-tracing simulation by the {\tt xissim} software. 
As a result, we find that the disagreements are 21\% and 11\% in FI and BI, respectively.
Extrapolating linearly these values to the off-angle position (\timeform{6'}),
we can expect the uncertainties are 36\% (FI) and 19\% (BI), which are consistent with our best-fit
results. Thus the second possibility is very likely.

We should note that any possible hard X-ray excess does not significantly change 
the best-fit thin thermal plasma model of the soft X-rays
within errors at the 90~\% confidence level apart from giving a lower plasma temperature 
by $\sim$ 0.2 keV. This however gives no essential change for the 
following discussions. 
 
\begin{table}[t]
\begin{center}
\caption{Best-fit parameters of G\,359.12$-$0.05}\label{table3}
 \begin{tabular}{ll}
  \hline
  \multicolumn{1}{c}{Parameter} & \multicolumn{1}{c}{Value} \\
  \hline
  $N_{\mathrm{H}}$ (10$^{22}$ cm$^{-2}$) &  6.7 (6.2--7.3) \\
  $kT$ (keV) & 0.93 (0.88--0.98) \\
  $Z_{\mathrm{Si}}$\footnotemark[$^*$] & 1.2 (0.85--1.7) \\
  $Z_{\mathrm{S}}$\footnotemark[$^*$] & 1.4 (1.2--1.7) \\
  $Z_{\mathrm{Ar}}$\footnotemark[$^*$] & 1.5 (1.1--2.1) \\
  Normalization\footnotemark[$\dagger$] & 6.4 (5.3--7.8) \\
  Flux\footnotemark[$\ddagger$] (photons~cm$^{-2}$~s$^{-1}$)& 6.3 $\times$ 10$^{-4}$ \\
  $\chi^2$/\textit{d.o.f} & 256/166 \\
  \hline
  \multicolumn{2}{@{}l@{}}{\hbox to 0pt{\parbox{75mm}{\footnotesize
        Notes. The errors are at the 90\% confidence level.
        \par\noindent
        \footnotemark[$^*$] Abundances in units of solar. The abundance of the other elements are fixed to the solar values.
        \par\noindent
        \footnotemark[$\dagger$] Normalization =  $(10^{-12}/4\pi D^2)\int n_{\mathrm{e}}n_{\mathrm{H}}dV$~cm$^{-5}$, 
where $D$, $n_{\mathrm{e}}$, and $n_{\mathrm{H}}$ are the distance to the G\,359.12$-$0.05, 
the electron density, and the hydrogen density, respectively. 
        \par\noindent
        \footnotemark[$\ddagger$] The observed flux of the plasma emission in the 1.0--10~keV band.
  }\hss}}
 \end{tabular}
\end{center}
\end{table}

\subsection{Spectrum of M\,359.23$-$0.04} 

\begin{figure}[t]
  \begin{center}
    \FigureFile(80mm,60mm){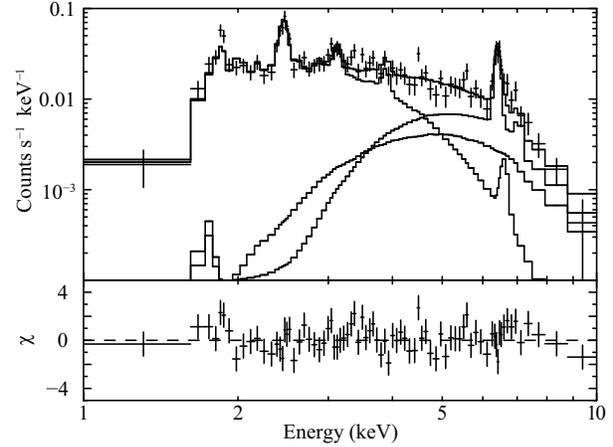}
  \end{center}
  \caption{The XIS background-subtracted FI spectrum of M\,359.23$-$0.04.
The solid lines are the best-fit models (see text).}\label{figure6}
\end{figure}

We extract the spectrum of M\,359.23$-$0.04 from the region designated in figure \ref{figure3}b.
The background region is the same as that for G\,359.12$-$0.05.
As shown in figure \ref{figure6}, the spectrum of M\,359.23$-$0.04 is similar to that of G\,359.12$-$0.05.
This is reasonable, because M\,359.23$-$0.04 is located in the diffuse emission of G\,359.12$-$0.05.
In addition, we find a strong line at 6.4~keV (Fe\emissiontype{I} K$\alpha$) and a hint of Fe\emissiontype{I} K$\beta$ 
at 7.1~keV.

We therefore  fit the spectrum with the same model of G\,359.12$-$0.05 adding an absorbed power-law plus 
two Gaussian lines at 6.4~keV and 7.1~keV.
The parameters of the plasma model 
are fixed at those of table \ref{table3} except the normalization.
The spill-over component of the GA is estimated with the same method as in the analysis of G\,359.12$-$0.05.  
The best-fit results and parameters are shown in figure \ref{figure6} and table \ref{table4}, respectively.
For visibility, only the FI spectrum is shown although the FI and BI spectra are fit simultaneously.

As the same reason of G\,359.12$-$0.05, the hard-band flux of M\,359.23$-$0.04  may have some ambiguity due to the 
uncertainty of the PSF. If we add the same amount of uncertainty (FI:31\% and BI:23\%) to the spill-over flux of the GA,  
the power-law continuum associated to  the 6.4-keV line decreases by $\sim~18$\%.  
This modifies the \textit{EW}$_\mathrm{6.4\,keV}$ value from 1.2~(0.83--1.4)~keV to 1.4~(1.1--1.7)~keV.

\begin{table}[t]
\begin{center}
\caption{Best-fit parameters of M\,359.23$-$0.04}\label{table4}
 \begin{tabular}{ll}
  \hline
  \multicolumn{1}{c}{Parameter} & \multicolumn{1}{c}{Value} \\
  \hline
  $N_{\rm{H}}$ (10$^{22}$ cm$^{-2}$) &  31 (19--35) \\
  Photon index & 2.9 (2.2--3.5) \vspace{1mm}\\
  Fe\emissiontype{I} K${\alpha}$:\\
  \phantom{0}Center (keV) & 6.41 (6.40--6.43) \\
  \phantom{0}Line flux\footnotemark[$*$] & 3.4 (2.6--4.0) \\
  \phantom{0}Equivalent width (keV) & 1.2 (0.83--1.4) \vspace{1mm}\\
  Fe\emissiontype{I} K${\beta}$:\\
  \phantom{0}Center (keV) & 7.06\footnotemark[$\dagger$] \\
  \phantom{0}Line flux\footnotemark[$*$] & $<$0.79 \\
  \phantom{0}Equivalent width (keV) & $<$0.36 \vspace{1mm}\\
  $\chi^2$/\textit{d.o.f} & 164/153 \\
  \hline
  \multicolumn{2}{@{}l@{}}{\hbox to 0pt{\parbox{85mm}{\footnotesize
	Notes. The errors are at the 90\% confidence level.
	\par\noindent
	\footnotemark[$*$] In units of 10$^{-5}$ photons~cm$^{-2}$~s$^{-1}$.
	\par\noindent
	\footnotemark[$\dagger$] Fixed to 1.1 $\times$ $E$ (Fe\emissiontype{I} K$\alpha$).
  }\hss}}
 \end{tabular}
\end{center}
\end{table} 

\section{Discussion}

\subsection{Nature of M\,359.23$-$0.04}

\begin{figure}[t]
  \begin{center}
    \FigureFile(80mm,120mm){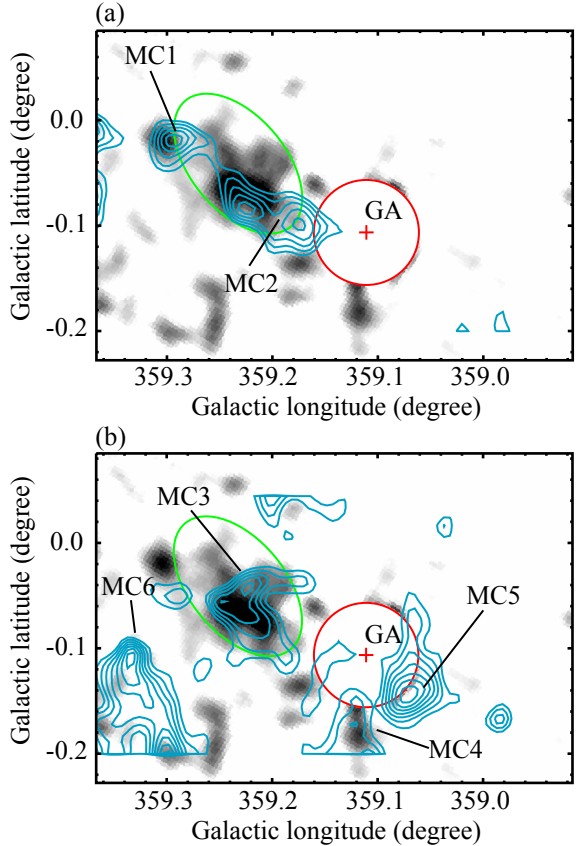}
  \end{center}
  \caption{The 6.4-keV line images (same as figure \ref{figure3}b) are shown in the gray scale. 
The blue contours trace the CS(\textit{J}=1-0) radio observation 
\citep{Tsuboi.1999} in the velocity bands of $-$140~to~$-$120~km~s$^{-1}$ (a), and in the $-$20~to~$-$0~km~s$^{-1}$ (b).
 The green ellipse shows the source region of M\,359.23$-$0.04. The red circle and cross indicate the \timeform{3'}-radius 
region and the position of the GA, respectively.
Individual prominent spots (clumps) in the radio contours are labeled as MC1 to MC6 (see text).}\label{figure7}
\end{figure}

From the radio MC observation of CS(\textit{J}=1--0) \citep{Tsuboi.1999}, 
we find that at least one clump of the MCs is associated with M\,359.23$-$0.04 
in the two velocity bands: the $-$140~to~$-$120~km~s$^{-1}$ band and $-$20~to~0~km~s$^{-1}$ band.
Figure \ref{figure7} shows the radio contours overlaid on the X-ray images. 
We find several clumps of MCs and designate them as MC1--6.
Among them, MC2 and MC3 are possibly associated with M\,359.23$-$0.04.

The best-fit \textit{EW}$_{\mathrm{6.4\,keV}}$ of M\,359.23$-$0.04 is 1.2--1.4~keV (see table \ref{table4} and section 3.5). 
This value is naturally explained by the X-ray irradiation (the X-ray reflection nebula: XRN model) 
on the MC with one solar abundance (e.g., \cite{Murakami.2000}; \cite{Nakajima.2009}).
If we adopt the electron bombarding model (\cite{Yusef-Zadeh.2007}; \cite{Fukuoka.2009}), 
we need more than 3 solar abundances.
The infrared observations (e.g., \cite{Davies.2009}) argued that the iron abundance of the GC is nearly equal to the solar value. 
In fact, most of the Suzaku observations of the GC region reported the abundances 
between 1--2 solar (e.g., \cite{Mori.2008}, \cite{Nobukawa.2008}; \cite{Tsuru.2009}; \cite{Nobukawa.2010}). 
Our observation also implies  1.2--1.5 solar abundances in this region
(see table \ref{table2} and \ref{table3}).
The XRN model is therefore more favorable for the M\,359.23$-$0.04 X-ray emission.

In the XRN model, the irradiating source should be very bright or be located near at M\,359.23$-$0.04. 
Thus the GA at a projected distance of \timeform{6'} from M\,359.23$-$0.04 and 
the past bright Sgr\,A$^*$ (e.g., \cite{Inui.2009}) are the possible 
candidates for the irradiating source. Another possible candidate is a transient source. 
Although this possibility due to an unknown transient is not 
excluded, the available observations place this possibility to be a less plausible hypothesis, 
because no bright transient source has been so far reported from 
the close vicinity of M\,359.23$-$0.04 (e.g.,\cite{in't Zand.2001}; \cite{Sakano.2002}; \cite{Muno.2009}).

Based on the observed profile (figure \ref{figure3}b),
the projected shape of M\,359.23$-$0.04 is approximated
with an ellipse of $12.5~\mathrm{pc} \times 22.5~\mathrm{pc}$ (minor $\times$ major axis).
Adopting a reasonable assumption that the 3-dimensional structure of M\,359.23$-$0.04 
is a prolate spheroid with the line-of-sight depth of 12.5~pc (the same as the minor axis), 
we estimate the volume to be $5 \times 10^{58}~\mathrm{cm}^{3}$ (a prolate spheroid with axises of $12.5~\mathrm{pc} \times 12.5~\mathrm{pc} \times 22.5~\mathrm{pc}$).
The observed \textit{N}$_{\mathrm{H}}$ is $(19\textrm{--}35)~\times 10^{22}$~cm$^{-2}$.
On the other hand, the past studies reported that the interstellar absorption to the GC region is \textit{N}$_{\mathrm{H}} = 6~\times 10^{22}$~cm$^{-2}$
(\cite{Sakano.2002}; \cite{Ryu.2009}).
To confirm this value, we conduct the model fitting of the GCDX in the close vicinity of the GA.
We then find that the interstellar absorption is \textit{N}$_{\mathrm{H}} = 5.7$~(5.5--6.0)~$\times 10^{22}$~cm$^{-2}$, 
which is consistent with the values in the references.
Subtracting the interstellar absorption to the GC region,
we estimate that the intrinsic \textit{N}$_{\mathrm{H}}$ is about $(1\textrm{--}3) \times 10^{23}$~cm$^{-2}$.
On the other hand, \textit{N}$_{\mathrm{H}}$ of MC2 or MC3, a counterpart of M\,359.23$-$0.04 is 
$\sim 1 \times 10^{23}~$cm$^{-2}$ from the CS(\textit{J}=1--0) observation \citep{Tsuboi.1999}.
Thus \textit{N}$_\mathrm{H}$ of the cloud is in the range of (1--3) $\times 10^{23}$~cm$^{-2}$.

Adopting these physical parameters and assuming that the distances of M\,359.23$-$0.04 from the GA 
and Sgr\,A$^*$ are the same as the projected distances,
we estimate the required luminosity of the GA and Sgr\,A$^*$ to produce the 6.4-keV-line flux
of M\,359.23$-$0.04. In this process, the spectrum shapes of the GA and Sgr\,A$^*$ are 
assumed to be a power-law with photon index  $\Gamma$ = 1.4 and 2, respectively
(see table \ref{table2} and \cite{Murakami.2000}).
The 6.4-keV-fluorescence line is due to X-rays above the iron K-edge energy at 7.1~keV \citep{Murakami.2000}.
At these high energy X-rays,
we can ignore the interstellar absorption between the irradiating source and the XRN.
The results of the required luminosity in the 2.5--12~keV band for the GA and Sgr\,A$^*$ 
are $4 \times 10^{36}~\mathrm{erg~s}^{-1}$ and 
$5 \times 10^{38}~\mathrm{erg~s}^{-1}$, respectively. 

The observed luminosity of the GA is smaller by a factor of 2 than the required luminosity 
(see table \ref{table2}). 
This difference is within the systematic errors of the size, density and distance;
i.e., the real distance would be larger than the projected distance.
In addition, the flux-variation of the GA  by a factor of 2--5 was reported  
in the long time period (\cite{Churazov.1993b}; \cite{Sakano.1999}; \cite{Smith.2002}). 
Thus the scenario of the GA origin is consistent with the observed results.

The other possibility, the past flare of Sgr\,A$^*$,  has been proposed 
for many other 6.4-keV clumps in the GC region (e.g., \cite{Nobukawa.2008}; \cite{Nakajima.2009}). 
The same argument can also be applied to M\,359.23$-$0.04, and thus the scenario of Sgr\,A$^*$ origin is possible.

In figure \ref{figure7}, we find other 6.4-keV clumps at near MC1 and MC4.
The 6.4-keV fluxes of them are $\sim$ 2 times fainter than that of M\,359.23$-$0.04, and hence
the physical association is less clear.
The other MCs, MC5 and MC6, show no association to the 6.4-keV emissions.
These features may not be surprising, if the putative past flare
was variable in the time scale of a few years \citep{Koyama.2009}.

On the other hand, the scenario of the GA origin can be more easily
assessed, if the real distances (not projected distances) of these clumps from the GA are   
1.4 times larger than those of MC2 or MC3, because the 6.4-keV fluxes from  MC1, MC4, MC5, and MC6 
are less than half of those from MC2 or MC3.

\subsection{Nature of the GA}

The observed column density of the GA is $\textit{N}_{\mathrm{H}} = 12 \times 10^{22}~\mathrm{cm}^{-2}$.
If the GA is associated with M\,359.23$-$0.04 (see section 4.1), 
the GA lies in the GC region, because \citet{Tsuboi.1999} showed that MCs of CS(\textit{J}=1--0) are in the GC region.
In this case, the large $\textit{N}_{\mathrm{H}}$ of the GA is explained by intrinsic absorption \citep{Kawai.1988} of 
the order of $10^{23}$~cm$^{-2}$.

We constrain the upper-limit for the \textit{EW}$_\mathrm{6.4\,keV}$ 
of 8~eV. 
The iron abundance is also accurately determined to be 1.38--1.44 solar. 
If the GA is fully covered by the gas for $\Omega$ = 4$\pi$ steradian, the upper-limit
of \textit{EW}$_\mathrm{6.4\,keV} = 8~\mathrm{eV}$ constrains the column density of the
circumstellar gas within 7.5~pc (\timeform{3'} at 8.5 kpc) to be
\textit{N}$_{\mathrm{H}}$ $\leq$ 4$\times$ 10$^{21}$~cm$^{-2}$ (\cite{Inoue.1985}). 
The implication of the small value of \textit{EW}$_\mathrm{6.4\,keV}$ was discussed
in detail by  \citet{Sakano.1999}.  Here we discuss briefly following \citet{Sakano.1999}, based on the more severe constraint of  
\textit{EW}$_\mathrm{6.4\,keV}$ (\textit{EW}$_\mathrm{6.4\,keV}$ = 15 eV in \cite{Sakano.1999} \textit{vs} 8 eV in our case).
The \textit{EW}$_\mathrm{6.4\,keV}$ of 8 eV implies that the local density near the GA is less than 
2 $\times$ 10$^{2}$~H~cm$^{-3}$.
From figure 6 of \citet{Sakano.1999}, the velocity of the GA must be less than 7~km~s$^{-1}$ 
under the Bondi-Hoyle accretion mechanism even if the GA has the 
mass of 20~\Mo. Since no data of the velocity dispersion for a black hole are 
available, we refer to that of the radio pulsars (neutron stars) from \citet{Faucher.2006}. We then conclude that
the possibility of the GA with a smaller velocity than 7~km~s$^{-1}$ is only $2 \times 10^{-3}$\%.  
Thus the Bondi-Hoyle accretion for the GA is extremely unlikely.

Although  the observed \textit{EW}$_\mathrm{6.4\,keV} < 8~\mathrm{eV}$ gives
the column density of the circumstellar gas within 7.5~pc to be 
\textit{N}$_{\mathrm{H}}$ $\leq$ 4$\times$ 10$^{21}$~cm$^{-2}$ (this value is
averaged over 4$\pi$ steradian), the intrinsic \textit{N}$_\mathrm{H}$ 
determined from the low energy absorption is in the order of $10^{23}$~cm$^{-2}$.
Thus the large circumstellar gas should be localized in front of the GA. 
\citet{Bally.1991} reported that an MC is associated to the GA, 
but no MC is found in the line of sight to the GA from \citet{Tsuboi.1999}.
This inconsistency may come from the poor spatial resolution of the former observation.
In fact, the spatial resolution of the former observation is $\sim$~\timeform{2'}, while the latter 
is $\sim$~\timeform{0.5'}.
Our implication is consistent with \citet{Tsuboi.1999}, and hence one possible scenario for 
the large absorption is due to an accreting gas; the GA is a binary system with a nearly edge-on 
accretion disk.

\subsection{Plasma Parameters of G\,359.12$-$0.05}

\begin{figure}[t]
  \begin{center}
    \FigureFile(80mm,65mm){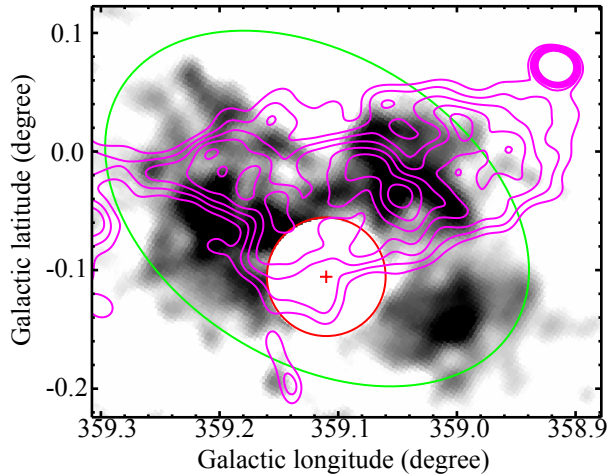}
  \end{center}
  \caption{The 2.45-keV line image (same as figure \ref{figure3}a) is shown in the gray scale. 
The magenta contours trace G\,359.07$-$0.02 from the 90~cm radio continuum observation \citep{LaRosa.2000}.
The green ellipse shows the source region of G\,359.12$-$0.05. The red circle and cross indicate the \timeform{3'}-radius 
region and the position of the GA, respectively.}\label{figure8}
\end{figure}

The observed column density of G\,359.12$-$0.05 $\textit{N}_{\rm{H}} = 6.7 \times 10^{22}~\mathrm{cm}^{-2}$ is the same as the GC typical 
column density of $N_{\rm{H}} = 6 \times 10^{22}~\mathrm{cm}^{-2}$, and hence G\,359.12$-$0.05 is likely to be in the GC region. 
Since the source area is $\sim$ $140~\mathrm{arcmin}^2 = 8.1 \times 10^{39}~\mathrm{cm}^{2}$ (at 8.5~kpc), 
the volume of the plasma is estimated as $\sim$ $(8.1~\times~10^{39}~\mathrm{cm}^2)^{\frac{3}{2}} = 7.3 \times 10^{59}~\mathrm{cm}^{3}$.
Using the filling factor of $f$, we estimate the electron density, total mass, and total thermal energy
to be  $n_\mathrm{e}$ $\sim$ 0.3~$f^{-\frac{1}{2}}$~cm$^{-3}$, $M \sim$ 2 $\times 10^{2}$~$f^{\frac{1}{2}}$~\Mo,
and $E_{\mathrm{th}}$ $\sim$ 1 $\times$ 10$^{51}$~$f^{\frac{1}{2}}$~erg, respectively. 

The estimated mass and thermal energy are slightly larger than those of the other SNRs in the GC region, 
but are consistent with those of a single SNR. 
As shown in figure \ref{figure8}, the radio SNR candidate G\,359.07$-$0.02 \citep{LaRosa.2000} coincides in position with 
G\,359.12$-$0.05. 
We therefore propose that G\,359.12$-$0.05 is an X-ray counterpart of G\,359.07$-$0.02.  The thin thermal plasma spectrum and the 
physical parameters place G\,359.12$-$0.05 to be an X-ray SNR in the GC region. 
From the size of $\sim$ $9.0~\times~10^{19}~\mathrm{cm}$ and the sound velocity of
$\sim$ $5~\times~10^{7}~\mathrm{cm}~\mathrm{s}^{-1}$ ($kT \sim0.9$ keV), the age of this SNR is estimated
to be $\sim$ $6~\times~10^{4}$ years. 

If the GA is also located in the GC region (see section 4.2), G\,359.12$-$0.05 may be physically associated with the GA. 
Then the GA would be a rare system which shows the association of a black hole and an SNR.
A similar system is SS433 and the radio 
SNR W50 (e.g., \cite{Safi-Harb.1997}).  The nature of SS433 is debatable but is a potential 
candidate of a black hole. On the other hand, the GA is a more established black-hole candidate.  
Thus our result provides a unique case of the association between a black hole and an X-ray SNR.

\section{Summary}
\begin{enumerate}
\item The equivalent width of the 6.4-keV line of the GA is 
most severely constrained.
\item We discover two diffuse emissions, G\,359.12$-$0.05 and M\,359.23$-$0.04, around the GA.
\item G\,359.12$-$0.05 is an X-ray counterpart of G\,359.07$-$0.02, and a middle-aged SNR in the GC region.
\item M\,359.23$-$0.04 is likely an XRN illuminated by either the GA or Sgr\,A$^*$. In the former case,
the GA is located in the GC region. This may suggest the association of a Galactic black hole with an SNR.
\end{enumerate}

\bigskip
The authors thank to all the Suzaku team members for their developing hardware and software, spacecraft operations, and instrument calibrations.
We also wish to thank M. Tsuboi at JAXA/ISAS for his useful comments on the radio observations of MCs.
This work is supported by the Grant-in-Aid for the Global COE Program 
"The Next Generation of Physics, Spun from Universality and Emergence", 
the Grant-in-Aid for Scientific Research A No.18204015 (KK), Scientific Research B 20340043 (TT), 
and Challenging Exploratory Research No.2054019 (KK),
all from the Ministry of Education, Culture, Sports, Science and
Technology (MEXT) of Japan. 
MN and HU are supported by JSPS Research Fellowship for Young Scientists.

\end{document}